**Strong many-body interactions in ultrathin anisotropic tin (II) monosulfide**

Abdus Salam Sarkar, Aamir Mushtaq, Dushyant Kushavah and Suman Kalyan Pal*

School of Basic Sciences, Indian Institute of Technology Mandi, Kamand, Mandi-175005, Himachal Pradesh, India.

Advanced Materials Research Centre, Indian Institute of Technology Mandi, Kamand, Mandi-175005, Himachal Pradesh, India.

Energy Center, Indian Institute of Technology Mandi, Kamand, Mandi-175005, Himachal Pradesh, India.

*Corresponding author

Email: suman@iitmandi.ac.in

**ABSTRACT**

Two-dimensional (2D) tin(II) monosulfide (SnS) with strong structural anisotropy has been proven to be a phosphorene analogue. However, difficulty in isolating very thin layer of SnS pose challenges in practical utilization. Here, we prepare ultrathin SnS via liquid phase exfoliation. With transmission electron microscopy, we identify the buckled structure of 2D SnS. We employ temperature dependent Raman spectroscopy to elucidate electron-phonon interactions, which reveals a linear phonon shifts. The active Raman modes of ultrathin SnS exhibit higher sensitivity to temperature than other 2D materials. Moreover, we demonstrate strong light-matter interaction in ultrathin SnS using Z-scan and ultrafast spectroscopy. Rich exciton-exciton and coherent exciton-photon interactions arising from many-particle excited effects in ultrathin SnS eventually enhances the nonlinear optical properties. Our findings highlight the prospects for the synthesis of ultrathin anisotropic SnS towards the betterment of thermoelectric and photonic devices.

**INTRODUCTION**

Successful isolation of 2D materials beyond graphene such as boron nitride (BN), transition metal dichalcogenides (TMDs) and phosphorene opens up a new horizon in material research[1-3]. The unique optical properties that arise with strong light-matter interaction at their 2D quantum limit can provide a plethora of opportunities for future energy conversion and photonic applications[4-7]. Of a particular note, the structural anisotropy of ultrathin 2D phosphorene has sparked the revival of fundamental research[8-10]. The intriguing in-plane anisotropy due to the difference in electronic structure along the armchair and zigzag directions gives rise to different refractive index along these directions[11]. Strong anisotropy can also lead to the formation of quasi excitons, biexcitons, trions and phonons[8-10,12]. In spite of a number of exotic phenomena, chemical stability of phosphorene is a critical issue[13-15].

2D SnS, a member of emerging layered metal monochalcogenide (MX, M = Ge, Sn etc.; X = chalcogen) family has recently been recognized as an analog of phosphorene[16]. Theoretical investigations revealed a sizeable band gap, odd-even quantum confinement effect, high carrier mobility and large absorption coefficient in few-layer SnS[17]. The first-principle simulations based on the modern theory of polarization predicted enormous, piezoelectric effect of monolayer SnS. One or two orders of magnitude larger value of the piezoelectric coefficient of SnS than other 2D materials was attributed to its unique puckered $C_{2v}$ symmetry and electronic structure[18]. The buckled layered structure of SnS (**Fig. 1a**) gives rise to the structural anisotropy, which is manifested in the Raman response[19], nonlinear optical property[20], electrical mobility[16] and photoactivity[21]. More importantly, SnS layers are environmentally, thermally and dynamically stable[17,22]. Previous reports on layered SnS are mainly theoretical, but experimental investigations are on the rise. O'Brien and coworkers[23] have reported exfoliation of thin SnS nanosheets containing 3-4 bi-layers. In the first report on light matter interaction in multilayer SnS sheets, Raman spectroscopic investigations have been carried out to know the thermal properties[19]. However, the revelation of ultrathin layer (monolayer/bi-layer) SnS nanosheets and their thermal properties are yet to be explored.

In a system of reduced dimensionality[24-27], exciton-exciton annihilation (EEA), a many body process in which one exciton disappears by donating its energy, is particularly significant. Recently, Nardeep et al.[28] noticed strong density-dependent initial decay of excitonic population in MoSe$_2$ monolayer which was well illustrated by EEA. The annihilation process was also apparent in monolayer $MoS_2$, mono-, bi- and tri-layer $WS_2$[27,29]. In layered $WS_2$, annihilation rate was found to be faster in monolayer than in bi- or tri-layer due to reduced many-body interaction and phonon-assisted annihilation of indirect excitons[29]. Nonetheless, 2D layered materials possess nonlinear optical properties that determine the performance of nanophotonic devices. Wang et al. observed much better saturable response in $MoS_2$ than graphene[30]. The broadband and enhanced saturable absorption response of multilayer black phosphorous (BP) could be used to develop broadband ultrafast mode-locker, passive Q-switcher and optical switcher[31-33]. Linear and nonlinear ultrafast absorption behaviors of chemically exfoliated phosphorene have also been investigated. Ultrashort pulse generation has been demonstrated by taking advantage of their unique nonlinear absorption[34]. However, further study of light matter interaction in anisotropic 2D materials is very important to exploit them in energy conversion and laser applications.

In this study, we report for the first time the preparation of ultrathin SnS nanosheets via liquid phase exfoliation. Cryogenic Raman spectroscopy, pump-probe spectroscopy and Z-scan techniques were employed to investigate the electron-phonon and exciton-exciton interactions in layered SnS sheets. A linear behavior of phonon energy shift with temperature is observed for ultrathin SnS. Our femtosecond transient absorption studies reveal strong quasiparticle interaction in ultrathin SnS leading to EEA, which in turn enhances nonlinear optical response by making them potential candidates for photonic applications.

## RESULTS AND DISCUSSION

**Ultrathin SnS nanosheets**

We conducted liquid phase exfoliation (LPE) of ultrathin buckled layer (**Fig. 1**) SnS nanosheets in acetone solvent (**see methods**) and characterized them with extinction spectroscopy and atomic force microscopy (AFM). The exfoliated SnS in acetone exhibits a wide absorption profile with a shoulder at ~420 nm (**see Supplementary Fig. 1a**). **Fig. 2a** depicts typical AFM images of SnS samples prepared via 20h of ultrasonication. The average sheet thickness of the sample is estimated to be 1.10 nm (**Fig. 2b**). The exfoliated sheets could be of few monolayer thick[35]. Sheet like structure of exfoliated SnS is evident from the TEM image (**Fig. 2c**). The crystal structure of layered SnS is orthorhombic with space group Pnma (**Fig. 1**)[36,37]. To determine the atomic structure of the exfoliated 2D SnS nanosheets, high resolution TEM (HRTEM) micrographs were captured (**Fig. 2d**). A perfect rhombus-like lattice fringes (**Fig. 2d**) with sharp selected area diffraction (SAED) pattern (**Fig. 2e**) confirms the single crystallinity of the ultrathin SnS. The corner angle (~85.6°) of an ultrathin SnS layer agrees well with the theoretical model shown in **Fig. 1b**. The associated Fourier transform of HRTEM image (**inset of Fig. 2f**) further supports the presence of orthorhombic phase in ultrathin SnS.

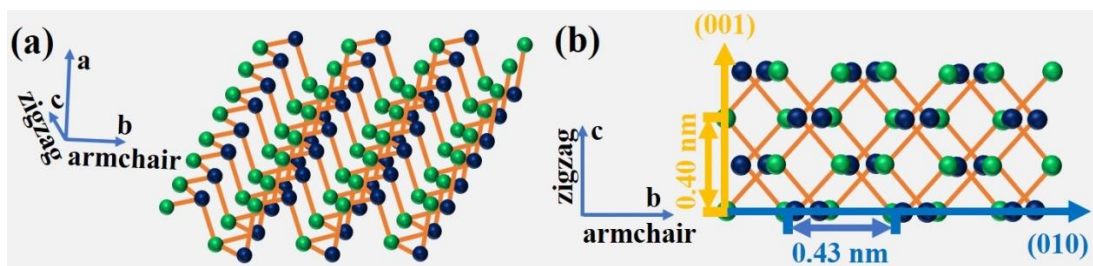

**Figure 1 | Model atomic structure of SnS crystal. a**, side view or 3D perspective and **b**, top view or along the interlayer direction (*b* and *c* axes represent the armchair and zigzag directions, respectively).

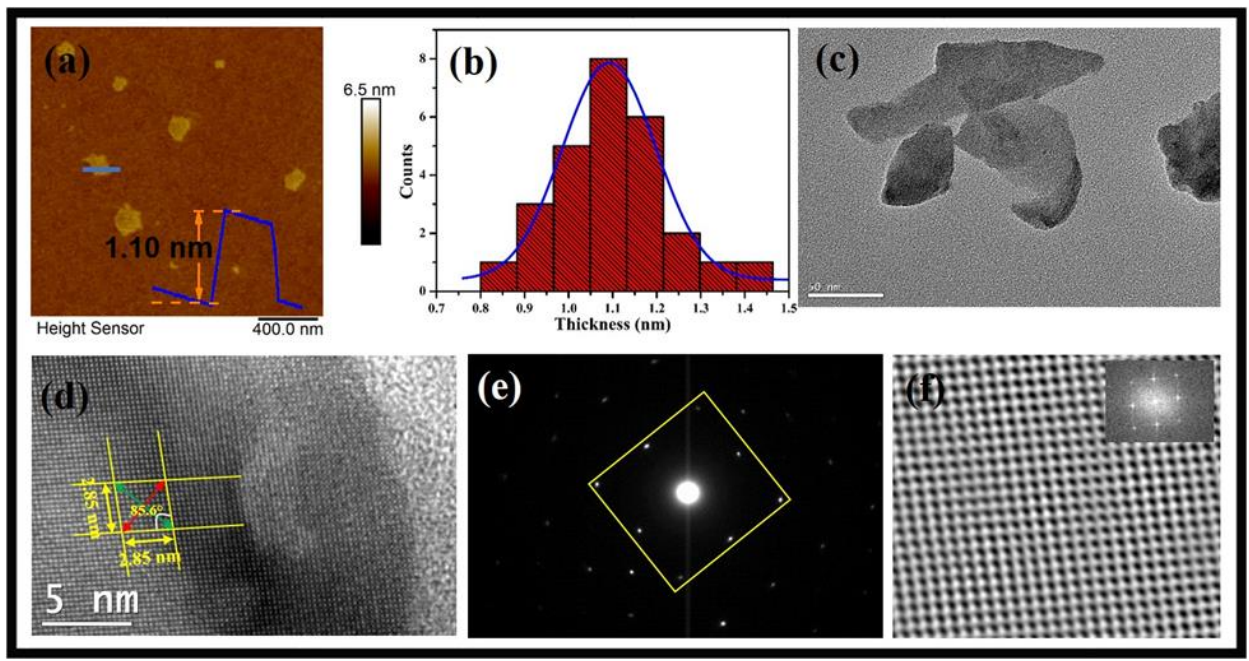

**Figure 2 | Microscopic characterization of layered SnS nanosheets. a,** Atomic force microscopy (AFM) image, **b,** Histograms of thickness distribution. Lateral dimension is ~170 nm (**see Supplementary Fig. 1b**). Transmission electron microscopy images. **c**, low resolution TEM, **d,** HRTEM image of ultrathin SnS. The length of lattice fringes (10 fringes) along two different directions are 2.85 and 2.85 nm, **e**, SAED pattern corresponding to the image in (d), **f**, FFT filtered atomic resolution of the selected area in (d) (Inset: FFT pattern of the selected region in (d)).

**Phonon dynamics**

Group IV layered monochalcogenide, SnS has a crystal structure that belongs to the orthorhombic crystal system with $D_{2h}$ symmetry (Pnma)[38,39]. In such structure, number of phonon modes (24 modes) at the center of the Brillouin zone can be expressed as $\Gamma = 4A_g + 2B_{1g} + 4B_{2g} + 2B_{3g} + 2A_u + 4B_{1u} + 2B_{2u} + 4B_{3u}$, where $A_g$, $B_{1g}$, $B_{2g}$, and $B_{3g}$ are Raman active modes[11,38]. **Fig. 3a** displays room temperature Raman spectra of bulk and exfoliated ultrathin nanosheets having thickness 1.10 nm. The inactive Raman modes ($A_g$ and $B_{3g}$ of the bulk) are found to be prominent while lowering the dimensionality (thickness) of the material. Unlike previous studies, four active Raman modes (at similar wave numbers) have been appeared in ultrathin SnS sheets[16,40-42]. Phonon peaks at 94.5, 161.1, and 221.5 cm$^{-1}$ are associated with the $A_g(1)$, $A_g(2)$, and $A_g(3)$ modes, respectively, while the peak at 167.7 cm$^{-1}$ is assigned to the $B_{3g}$ mode (**Fig. 3b**). The transformation of bulk SnS to ultrathin leads to a significant shift (3.46 cm$^{-1}$) in $A_g(1)$ mode, which could be attributed to the evolution of coupling between electronic transitions and phonons[43].

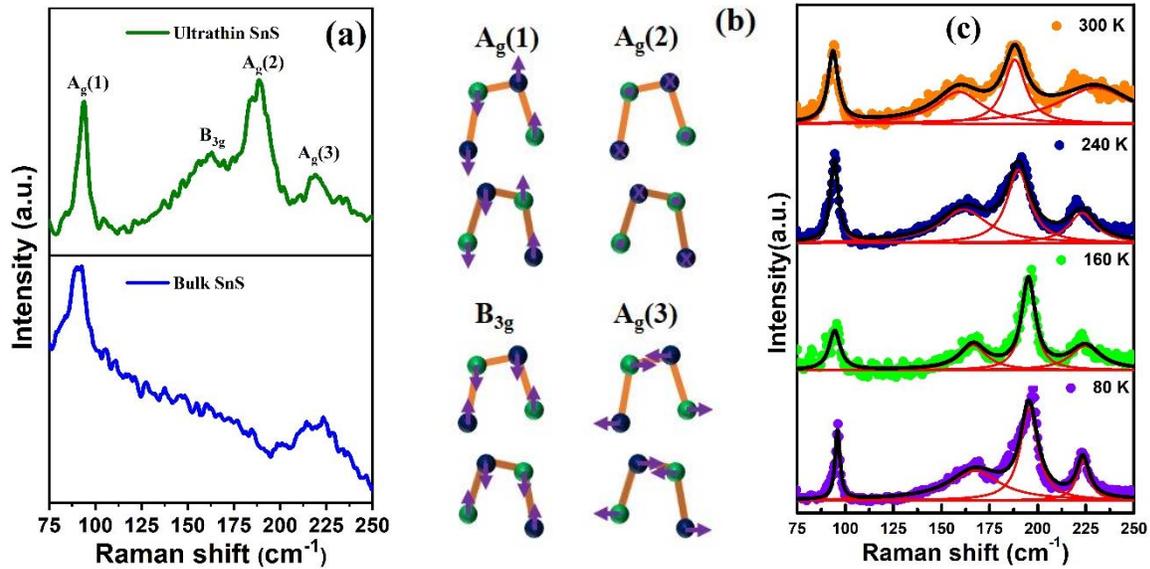

**Figure 3 | Raman spectra of SnS. a**, Room temperature Raman spectra of bulk and ultrathin nanosheets prepared by LPE. **b**, Atomic displacements corresponding to different Raman active modes of layered SnS (green and blue dots represent sulfur and tin atoms, respectively). **c**,

Raman shift of various vibrational modes of SnS nanosheets at different temperatures (incident laser wavelength and power density are 532 nm and 3.21 mW/μm$^2$, respectively).

Further, we focus on the phonon dynamics in ultrathin SnS. The cryogenic Raman spectra of SnS sheets at selected temperatures (between 80 to 300 K) are presented in **Fig. 3c**. A significant downshift of phonon peak position (associated to the phonon energy, ℏω) and broadening (related to the phonon lifetime) of Raman modes at higher temperatures are clearly evident (**see Supplementary Information Fig. 2**). All active modes in Raman spectra were fitted (**Fig. 3c**) to acquire quantitative information. The calculated peak position and linewidth (full width at half maximum, FWHM) of the Raman modes are plotted in **Fig. 4**. The evident phonon energy shift and lifetime variation may arise from thermal expansion, electron-phonon and anharmonic phonon-phonon interactions[44]. However, the positive temperature dependence of linewidth may be associated with optical phonon decay[45,46].

All the phonon modes of exfoliated SnS vary linearly with temperature (**Fig. 4**). Similar linear variation of phonon modes has already been reported in other anisotropic layered materials[8]. Furthermore, to describe the decrease in phonon energy in layered SnS with temperature change, the Grüneisen model (equation 1) was used[47,48].

$$\omega(T) = \omega_0 + \chi(T) \ldots\ldots\ldots\ldots\ldots\ldots\ldots\ldots(1)$$

where, $\omega_0$ is the frequency of the vibration at the absolute zero of temperature and $\chi$ is the first-order temperature coefficient (TC) of a Raman mode. Values of first order TCs for each modes of layered SnS were obtained by fitting experimental data (**Fig. 4 (a-d)**). The calculated $\chi$ values are compared with the previously reported values for SnS and other 2D materials (**see Supplementary Table 1**). In general, the TC of each Raman active modes of a layered 2D material is crucial in thermal metrology based on the cryogenic Raman spectroscopy. It helps to find the thermal properties of the material and to examine superiority of a material for energy conversion and electronic device applications[19]. Moreover, the energy conversion efficiency of a thermoelectric material can directly be improved by suppressing the propagation of phonon responsible for macroscopic thermal transport. In this scenario, the relatively high values of $\chi$ make ultrathin SnS sheet a superior (than other contemporary 2D materials) building block for thermoelectrics.

Nonetheless, phonon linewidth in ultrathin (**Fig. 4e-h**) SnS is temperature dependent. The change in phonon linewidth with increasing temperature is small in layered SnS. This behavior in layered SnS is observed due to a double resonance process, which is active only in a single and/or a few atomic layer thick nanosheets. To understand the origin of phonon linewidth in ultrathin 2D materials, the knowledge of phonon dispersion is essential. Many body calculations for layered materials having several phonon modes suggest that linewidth arises from the decay of the zone center optical phonon into one optical and one acoustic phonon belonging to the same branch[49,50]. The broadening of a Raman line can be expressed as[49]

$$\Gamma(T) = \Gamma_0 + A[1 + n(\omega_1, T) + n(\omega_2, T)]\ldots\ldots\ldots\ldots\ldots(2)$$

where, $\Gamma_0$ is the background contribution, $A$ is anharmonic coefficient, and $n(\omega, T)$ is the Bose-Einstein distribution function.

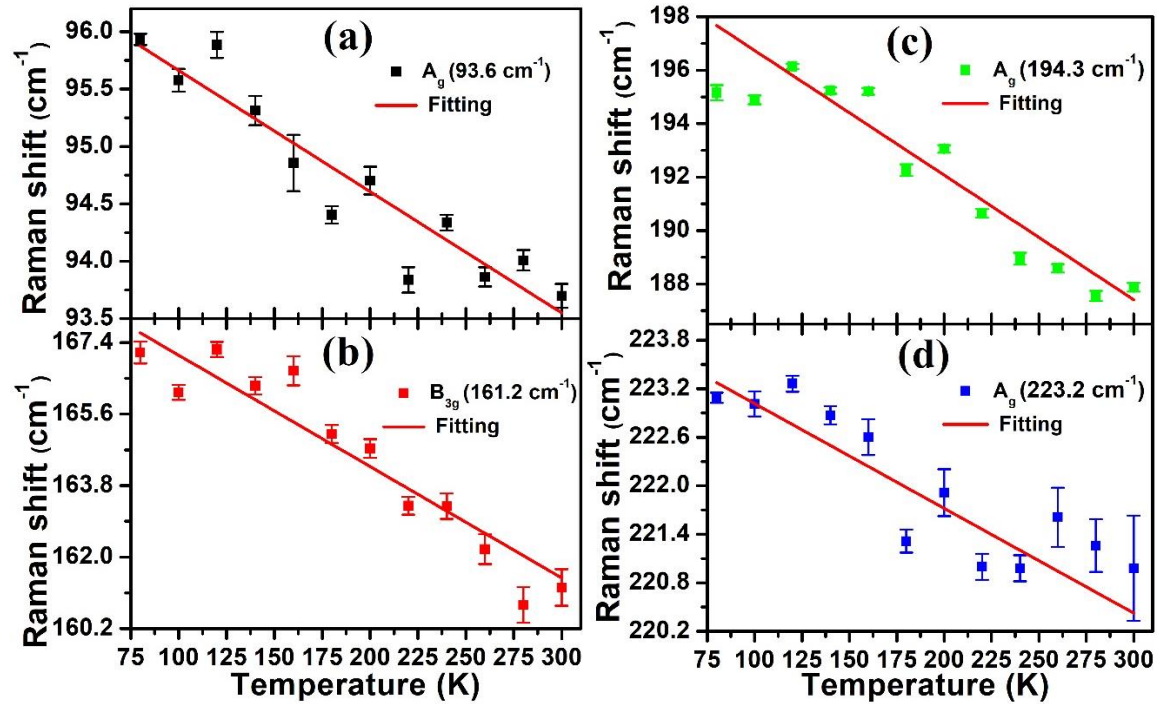

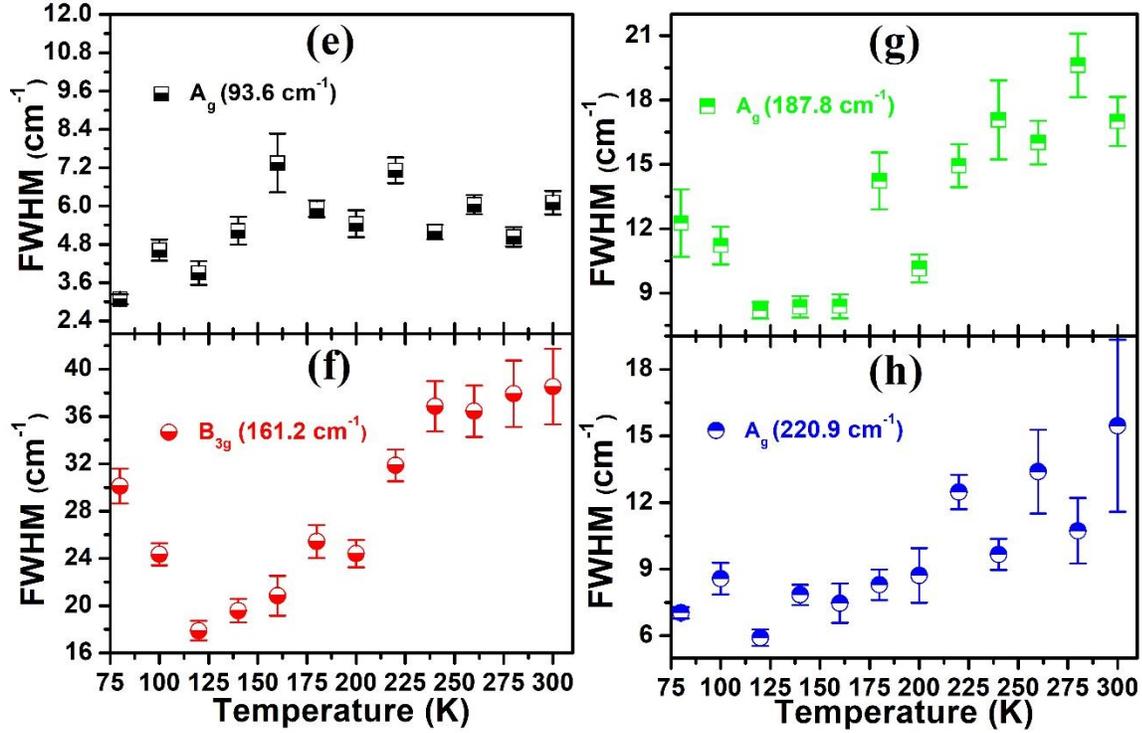

**Figure 4 | Temperature dependent phonon frequencies and linewidth in ultrathin SnS.** Temperature dependence of the phonon frequency (active $A_g$ and $B_{3g}$ phonon modes) (**a-d**). The solid squares are experimental data points and solid lines represent fit to Grüneisen model[47]. Temperature dependence of phonon linewidth (active $A_g$ and $B_{3g}$ modes) (**e, h**). The error bars represent the average uncertainty in determining peak position or FWHM.

**Saturable absorption**

In our investigation of NLO properties of layered SnS, we used femtosecond laser based Z-scan technique **(Fig. 5a and methods)**[51]. Typical open aperture (OA) Z-scan curve for ultrathin SnS nanosheets obtained at a peak intensity of 213 GW/cm$^2$ is presented in **Fig. 5b**. The normalized transmittance is increased to a maximum as the sample reaches the focus (Z=0). This is the consequence of saturable absorption, which eventually leads to the increase in total transmittance as the intensity of the incident beam is increased. Since the band gap of SnS nanosheets (1.6 eV) is slightly larger than the energy of excitation photon (1.55 eV), the saturable absorption response arises from optical transitions and band filling of edge states[52]. On the other hand, absence of any signal in the Z–scan measurement for pure acetone **(see Supplementary Fig. 3)** ensures that saturable absorption is solely the property of layered SnS. Further, for quantitative understanding of the NLO response, the Z-scan measurement data were fitted by the equation **(Fig. 5b)**[51]

$$T = 1 - \frac{\alpha_{NL} I_0 L_{eff}}{2\sqrt{2}\left[1+\left(\frac{Z}{Z_0}\right)^2\right]} \quad \ldots \ldots \ldots \ldots \quad (3)$$

$\alpha_{NL}$ is NLO absorption coefficient, $I_0$ is on-axis peak intensity, $Z_0$ is the diffraction length of the beam, $L_{eff} = \frac{(1-e^{-\alpha_0 L})}{\alpha_0 L}$, effective path length, $\alpha_0$ is linear absorption coefficient and L is the sample path length. The value of $\alpha_{NL}$ is found to be -0.5 ×10$^{-3}$ cm/GW. The value of saturable intensity was obtained by fitting the transmittance data **(Fig. 5c)** with the relation[53]

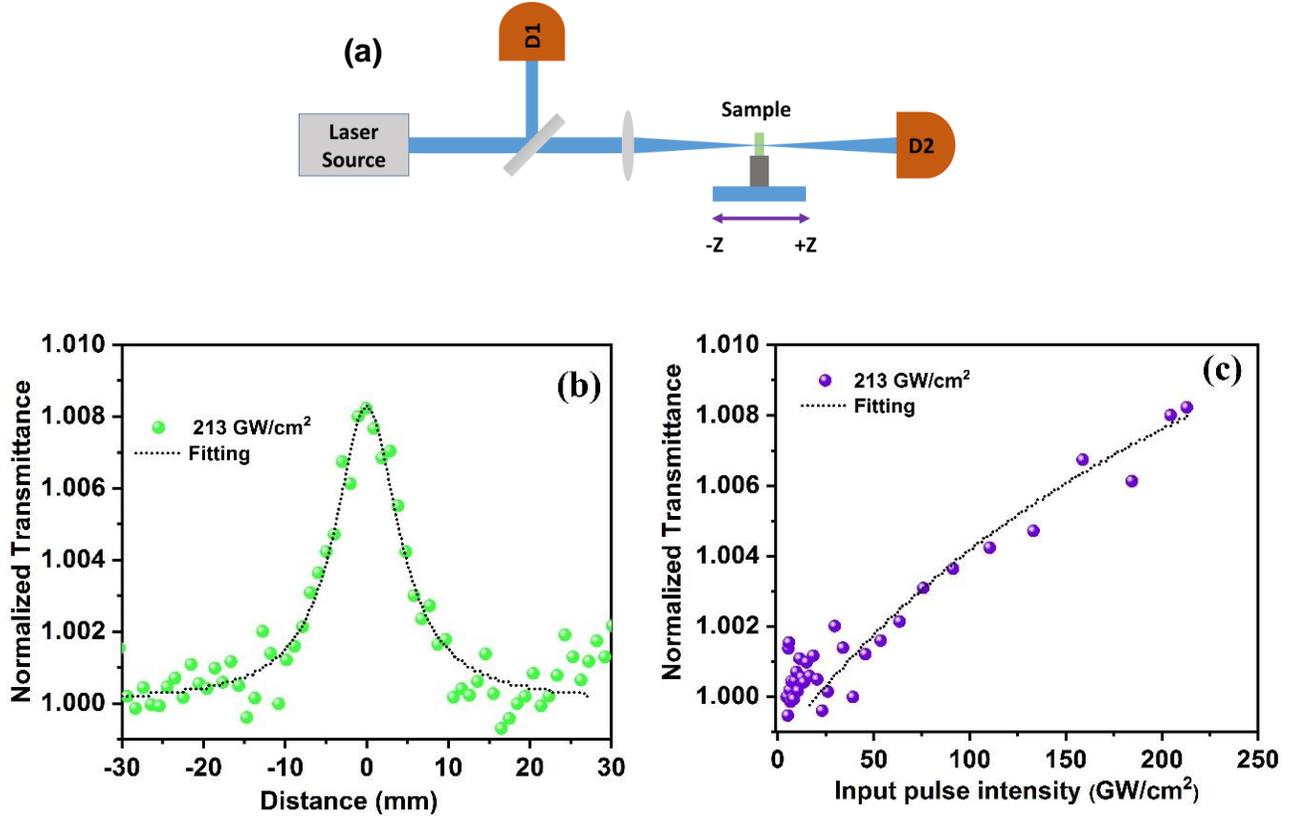

**Figure 5 | Z-scan results of the ultrathin SnS nanosheets for the fs pulses at 800 nm. a,** Schematic sketch of Z-scan technique. **b,** Open aperture (OA) Z-scan curve. **c,** Normalized transmittance as a function of input pulse intensity.

$$T = 1 - \frac{\alpha_s}{1+\frac{I_z}{I_{sat}}} - \alpha_{ns} \dots\dots\dots\dots\dots\dots\dots\dots\dots\dots\dots\dots(4)$$

where $\alpha_s$ is modulation depth, $I_{sat}$ is saturable intensity, $I_z = \frac{I_0}{1+\frac{Z^2}{Z_0^2}}$ is input intensity and $\alpha_{ns}$ is nonsaturable loss. The estimated value of $I_{sat}$ is 570 GW/cm$^2$. The NLO properties of SnS nanosheets are tabulated with other 2D materials which have shown promise as saturable absorber (**see Supplementary Table 2**). The observed high nonlinear absorption coefficient and low saturable intensity of ultrathin SnS similar to these 2D materials evince its great potential as a saturable absorber in ultrafast photonics devices.

**Exciton dynamics**

We employed ultrafast transient absorption spectroscopy to study the kinetics of excitons and their interactions. **Fig. 6a** shows TA spectra of SnS nanosheets obtained after excitation with a pump tuned at 480 nm, where two bands can be seen at 550 and 680 nm. As long as excitation intensity is low, the rise of high energy band (580 nm) is faster than the low energy band (680 nm). Strikingly, at higher pump intensity, the TA spectra are dominated by the low energy band (**Fig. 6b**).

We first focus our attention to find the origin of the TA band at 550 nm at low excitation intensity. TA kinetics were recorded at 550 nm using 480 nm pump at four representative pump fluences **as shown in Fig. 6c.** TA signals undergo a slow decay at longer time followed by fast early time decay. The closer view of signals (**see Supplementary Fig. 4a**) indicates that TA signal at early time is highly dependent on the incident photon density. The observed variation of peak TA signal with incident photon density can precisely be recounted by a saturation absorption model[28]

$$\Delta A \propto \frac{N}{N_s+N} \dots\dots\dots\dots\dots\dots\dots\dots\dots\dots(5)$$

where $N$ and $N_s$ are exciton density and saturation exciton density, respectively. The fitting of experimental data (**see Supplementary Fig. 4b**) yields $N_s = 1.4 \times 10^{13}$ photons/cm$^2$. We must mention that this saturation exciton density is consistent with that obtained from Z-scan measurement. To know the effect of exciton density on the slow decay process, we fitted the data after 10 ps to a single exponential function (**see Supplementary Fig. 4c**). The decay time (obtained from fitting) scales linearly with the pump fluence (**see Supplementary Fig. 4d**)

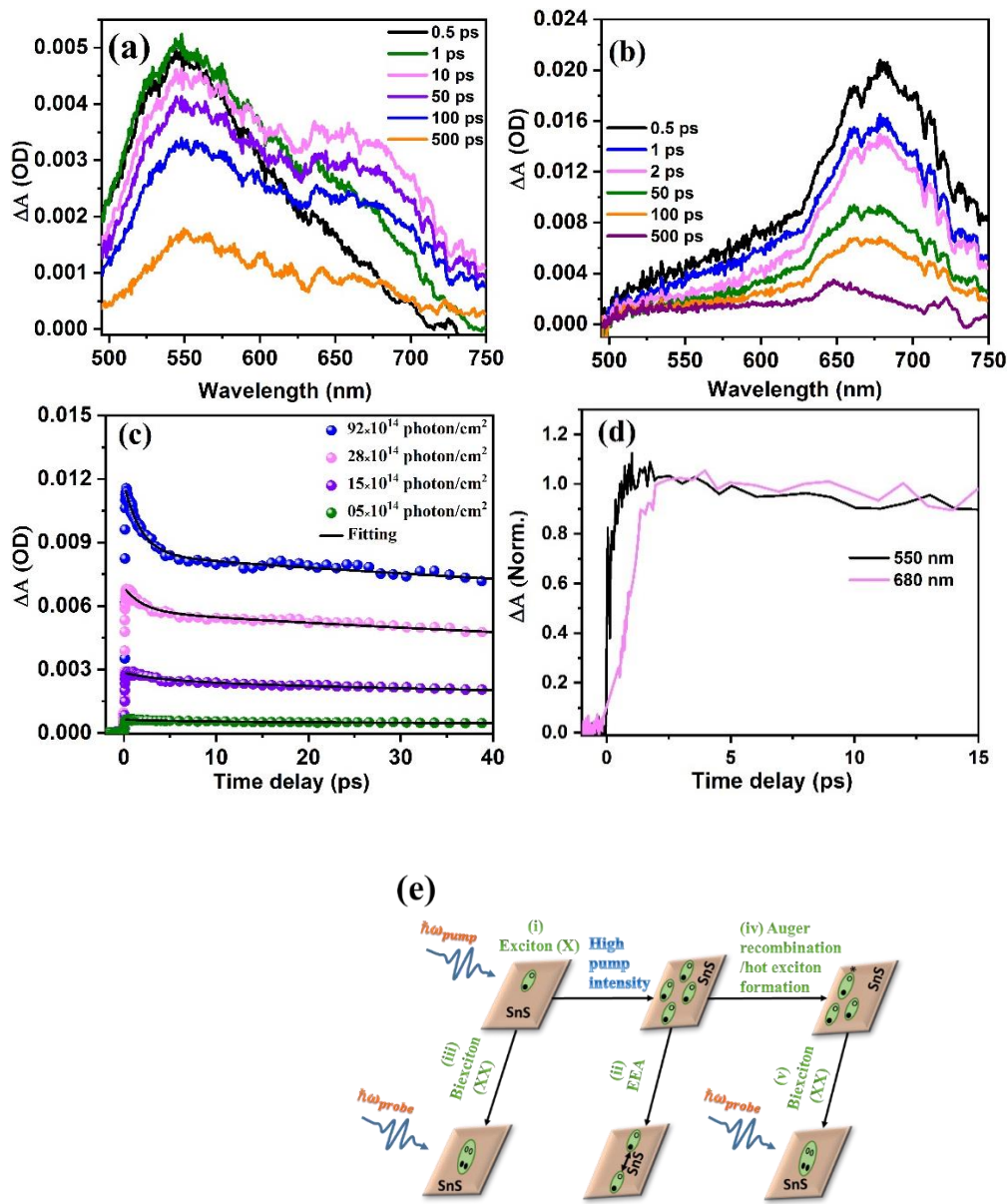

**Figure 6 | Transient absorption (TA) study of ultrathin SnS nanosheets.** TA spectra at pump fluence: **a,** $15 \times 10^{14}$ photons/cm$^2$, **b,** $28 \times 10^{14}$ photons/cm$^2$. Samples were excited with 480 nm laser beam. **c,** Exciton dynamics at different pump fluences (pump and probe beams were tuned at 480 and 550 nm, respectively). **d,** Normalized TA kinetics at 550 (black) and 680 (pink) nm after exciting with a 480 nm laser of fluence $28 \times 10^{14}$ photons/cm$^2$. **e,** Summary of different processes involving multicarrier states in SnS nanosheets: (i) Exciton generation by

pump photon, (ii) EEA due to exciton-exciton interaction, (iii) Biexciton formation from an exciton by absorbing one high energy probe photon, (iv) Hot exciton generation via Auger process. (v) Biexciton formation from a hot exciton by absorbing one low energy probe photon.

suggesting that the related process is independent of excitation intensity. Next, we contemplate the pump fluence dependent decay (rate increases with pump intensity) of TA signals at early probe delays. The rate of annihilation of excitons is proportional to the square of the exciton density, $N_x$[27].

$$\frac{dN_x}{dt} = -kN_x^2 \quad \ldots\ldots\ldots\ldots\ldots\ldots\ldots\ldots\ldots\ldots(6)$$

where $k$ is the annihilation rate. The solution of the differential equation (6) is of the form,

$$N_x(t) = \frac{N_0}{1+kN_0 t} + C \quad \ldots\ldots\ldots\ldots\ldots\ldots\ldots\ldots\ldots(7)$$

Here, $N_0$ and C are the initial and residual exciton populations, respectively. The TA data at different excitation intensities are fitted well with equation (7) (**Fig. 6c**) suggesting a close link between the early time TA signal and the excitonic population, which can undergo annihilation at high exciton density. However, the signal observed in our TA experiments is an induced absorption, whereas the signature of exciton would appear as an induced transmission (bleaching). Thus, TA at early time cannot be the fingerprint of single exciton. A more compelling interpretation for positive TA feature at 550 nm is the creation of a biexciton (*XX*) by the probe pulse, from the single exciton (*X*), which was created by the pump pulse (*X* → *XX* transition). The dynamics of such biexciton is similar to that of exciton. Nonetheless, it should be noted that excitation of nearby molecules undergo annihilation in strongly confined system like organic crystals[54,55]. Strong Coulomb interactions between carriers not only lead to many-body bound state (e.g., biexciton, trion) formation but also result in highly efficient Auger processes such as EEA in 1-D (carbon nanotubes, Graphene nanoribbons) and 2D TMDs ($MoS_2$, $MoSe_2$ etc). Hence, the observed decay that varies quadratically with exciton density in ultrathin SnS could be attributed to exciton-exciton annihilation. The average rate of annihilation is determined to be $4.34 \times 10^{-4}$ $cm^2$/s at room temperature.

So far our discussion was limited to the origin of the TA band at 550 nm. We discern that this band is almost evanesced at higher excitation intensities followed by a concomitant rise of another band at 680 nm. Although, the low energy TA band appears late, the decay dynamics of both the band is same (**Fig. 6d**). These observations infer that the final state is same (biexcitonic) for both the transitions, but the TA band at 680 nm may develop from different initial state. At high laser fluence, Auger process could lead to the formation of trion, which lives longer than biexciton[56]. However, as the decay time of the 680 nm band is same as the biexcitonic band, we ruled out the possibility of trion formation and speculate the formation of biexciton from high energy exciton, which is generated via Auger process. Finally, to summarize the whole process (**Fig. 6e**), absorption of the pump pulse by SnS sheets creates exciton, which form biexciton by absorbing a probe photon. The excitonic population decays quickly via EEA at high exciton density. However, at higher laser fluence, inter excitonic interaction leads predominantly to Auger recombination, rather than EEA resulting in the formation of hot exciton, which further absorb a low energy probe photon and results in another biexciton.

**CONCLUSION**

We successfully conduct liquid phase exfoliation of SnS to obtain ultrathin (~ 1.10 nm) nanosheets. Such layered SnS exhibits a linear phonon dynamics (or phonon shift). Raman modes of 2D SnS are much more sensitivity to temperature than other layered materials. Furthermore, ultrathin SnS possesses strong nonlinear optical behavior leading to saturable absorption. The strong Coulomb interactions give rise to EEA (rate ~ $4.34 \times 10^{-4} cm^2/s$) resulting in a fast decay of the exciton population. However, at higher exciton density, Auger process leads to the formation of hot exciton, which is converted to biexciton by absorbing another photon of low energy. High nonlinear absorption coefficient and low saturable intensity of ultrathin SnS nanosheets due to strong many-body interactions make them suitable for ultrafast Q-switching. This work sheds light on the nature of the phonon dynamics and nonlinear optical phenomena in anisotropic 2D SnS for futuristic energy conversion and photonic applications.

## METHODS

**Liquid phase exfoliation (LPE) of ultrathin SnS nanosheets**

Tin (II) sulfide granular (>99.99%) trace metals basis was purchased from Sigma Aldrich and used without further purification. HPLC grade acetone was received from Merck. Tin (II) sulfide (3 mg/ml) was dissolved in acetone in nitrogen flushed glass vials and sealed with teflon-tape (m-seal) inside the nitrogen-filled globe box (Lab Star, MBraun, Germany) with $O_2$ and $H_2O$ levels <0.5 ppm. The sealed vials (solution) were ultrasonicated in a bath sonicator (LMUC-4 from Spectrochrome Instruments, India) for 20 h under 100 W power and 40 kHz frequency. Room temperature was maintained throughout the sonication process. After ultrasonication, a dark brown solution was obtained. To isolate SnS sheets, dispersed solutions were centrifuged (Tarson Product Pvt. Ltd.) at 8000 rpm for 15 min. The suspensions were then pipetted to obtain layered SnS sheets, which were used for further experiments.

**Extinction spectroscopy**

The UV-vis extinction spectra were recorded using Shimadzu UV-2450 spectrometer (Agilent Technologies, USA). All measurements were carried out with dispersion of exfoliated SnS nanosheets in solvent using a quartz cuvette (path length of 10 mm).

**Atomic force microscopy (AFM)**

Tapping mode AFM images were acquired in an instrument from Dimension Icon with ScanAsyst, Bruker, USA. The samples were prepared by dropcasting dispersion of layered SnS on cleaned Si substrate. The samples prepared from SnS nanosheets dispersed in acetone were dried at 100°C for 20 min. The annealing (solvent evaporation) was performed inside a glove box (Lab Star, MBraun, Germany) with $O_2$ and $H_2O$ levels <0.5 ppm. The annealed sample was placed on the AFM stage to scan. Pertinent scanning parameters were as follows: scan rate for all measurements - (0.678 Hz for Fig. 2a), (0.565 Hz for Fig. 2b) and (0.628 Hz for Fig. 2c); aspect ratio: 1:1; resolution: 256 samples/line, 256 lines.

**Transmission electron microscopy (TEM)**

High resolution TEM (HRTEM) analysis of the samples was performed by a TECNAI G2 200 kV (FEI, Electron Optics) electron microscope with 200 kV input voltage. Exfoliated SnS samples were dropcasted onto 300 square mesh cupper grids (from ICON) covered with a carbon film. The dropcasted grids were dried inside the glove box ($O_2$<0.5 ppm and $H_2O$<0.5 ppm) at room temperature. Further, precautions were taken to avoid dust particles.

**Raman spectroscopy**

Raman spectroscopic measurements were carried out in the backscattering geometry using a confocal microRaman spectrometer (LABRAM HR Evolution, Horiba JobinYvon SAS) coupled with a Peltier cooled CCD and spectrometer. An unpolarized laser source of wavelength 532 nm was used to excite the samples. The incident laser beam was focused using a 50X long-working distance objective with a numerical aperture, NA = 0.05 and 1800 g/mm grating on the sample. The incident laser power density was 3.21 mW/$\mu m^2$. For low temperature Raman spectroscopy measurements, the samples were placed inside a continuous flow liquid nitrogen cryostat (Linkam Scientific) with a stability of 0.1 K. The samples were prepared by dropcasting dispersion of SnS sheets on cleaned Si substrates and drying them inside the glove box ($O_2$<0.5 ppm and $H_2O$<0.5 ppm) at room temperature. The collected spectra were fitted with Lorentzian functions in OriginPro8.5.

**Z-scan**

The Z-scan setup is consist of a Ti:sapphire regenerative amplifier (Spitfire ace, Spectra Physics) seeded by an oscillator (Mai Tai SP, Spectra Physics) and two photodetectors D1 and D2 from Pascher instrument as shown in Fig. 5a. The sample is mounted over the computer controlled translational stage, which moves with 1 mm steps. Incident beam is splitted into two beams, namely reference and transmitted beam. Reference beam goes into D1 whereas second beam is focused down by a lens (focal length, 10 cm) over the sample and finally received by D2. Various filters have been used to minimize the intensity to avoid bubble formation and white light generation in the sample. A laser light of wavelength 800 nm, pulse width 57 fs and repetition rate 1 kHz was used for measurements. Dispersion of ultrathin SnS in acetone was taken in a 2 mm cuvette for the open aperture Z-scan measurements.

**Transient absorption**

Ultrafast transient absorption (TA) was measured using a pump-probe set up, which is analogous to one reported by us[6,57,58]. In short, a Ti:sapphire regenerative amplifier (Spitfire ace, Spectra Physics) seeded by an oscillator (Mai Tai SP, Spectra Physics) was used as a light source. Two beams of pump and probe pulses have been generated from the amplifier output having wavelength 800 nm, pulse energy 4mJ and pulse width < 35 fs. White light continuum (WLC) probe has been generated by sending fraction of 800 nm light through a focusing lens to a sapphire crystal. Resulting beam was then splitted into two beams, namely reference and sample, and detected separately. Pump blocked and unblocked conditions were created for detection with the help of a mechanical chopper operating at 500 Hz by blocking every alternate pulse. TA spectra were recorded by CCD arrays after dispersion through a grating spectrograph (Acton spectra Pro SP 2300). TA kinetics was recorded by two photodetectors from Pascher Instrument. All TA kinetic traces were fitted using MATLAB and Origin software.

**ACKNOWLEDGEMENTS**

The authors acknowledge Science and Engineering Research Board (SERB), Government of India (Grant No. SB/S1/OC-48/2014), for the financial support. Advanced Materials Research Center facilities of IIT Mandi are also being acknowledged.

**REFERENCES**

1      Manzeli, S., Ovchinnikov, D., Pasquier, D., Yazyev, O. V. & Kis, A. 2D transition metal dichalcogenides. *Nat. Rev. Mater.* **2**, 17033 (2017).

2       Akhtar, M. *et al.* Recent advances in synthesis, properties, and applications of phosphorene. *npj 2D Mater. Appl.* **1**, 5 (2017).

3       Zhang, K., Feng, Y., Wang, F., Yang, Z. & Wang, J. Two dimensional hexagonal boron nitride (2D-hBN): synthesis, properties and applications. *J. Mater. Chem. C* **5**, 11992-12022 (2017).

4       Seixas, L., Rodin, A., Carvalho, A. & Neto, A. C. Exciton binding energies and luminescence of phosphorene under pressure. *Phys. Rev. B* **91**, 115437 (2015).

5       Berkelbach, T. C., Hybertsen, M. S. & Reichman, D. R. Theory of neutral and charged excitons in monolayer transition metal dichalcogenides. *Phys. Rev. B* **88**, 045318 (2013).

6       Mushtaq, A., Ghosh, S., Sarkar, A. S. & Pal, S. K. Multiple Exciton Harvesting at Zero-Dimensional/Two-Dimensional Heterostructures. *ACS Energy Lett.* **2**, 1879-1885 (2017).

7       Sarkar, A. S. & Pal, S. K. A van der Waals p–n Heterojunction Based on Polymer-2D Layered MoS$_2$ for Solution Processable Electronics. *J. Phys. Chem. C* **121**, 21945-21954 (2017).

8       Zhang, S. *et al.* Extraordinary photoluminescence and strong temperature/angle-dependent Raman responses in few-layer phosphorene. *ACS Nano* **8**, 9590-9596 (2014).

9       Xu, R. *et al.* Extraordinarily bound quasi-one-dimensional trions in two-dimensional phosphorene atomic semiconductors. *ACS Nano* **10**, 2046-2053 (2016).

10      Yang, J. *et al.* Optical tuning of exciton and trion emissions in monolayer phosphorene. *Light Sci. Appl.* **4**, e312 (2015).

11      Zhang, S. *et al.* Spotting the differences in two-dimensional materials–the Raman scattering perspective. *Chem. Soc. Rev.* **47**, 3217-3240 (2018).

12  Miao, X., Zhang, G., Wang, F., Yan, H. & Ji, M. Layer-Dependent Ultrafast Carrier and Coherent Phonon Dynamics in Black Phosphorus. *Nano Lett.* **18**, 3053-3059 (2018).

13  Favron, A. *et al.* Photooxidation and quantum confinement effects in exfoliated black phosphorus. *Nat. Mater* **14**, 826 (2015).

14  Wood, J. D. *et al.* Effective passivation of exfoliated black phosphorus transistors against ambient degradation. *Nano Lett.* **14**, 6964-6970 (2014).

15  Ahmed, T. *et al.* Degradation of black phosphorus is contingent on UV–blue light exposure. *npj 2D Mater. Appl.* **1**, 18 (2017).

16  Tian, Z., Guo, C., Zhao, M., Li, R. & Xue, J. Two-dimensional SnS: A phosphorene analogue with strong in-plane electronic anisotropy. *ACS Nano* **11**, 2219-2226 (2017).

17  Xin, C. *et al.* Few-layer tin sulfide: a new black-phosphorus-analogue 2D material with a sizeable band gap, odd–even quantum confinement effect, and high carrier mobility. *J. Phys. Chem. C* **120**, 22663-22669 (2016).

18  Fei, R., Li, W., Li, J. & Yang, L. Giant piezoelectricity of monolayer group IV monochalcogenides: SnSe, SnS, GeSe, and GeS. *Appl. Phys. Lett.* **107**, 173104 (2015).

19  Xia, J. *et al.* Physical vapor deposition synthesis of two-dimensional orthorhombic SnS flakes with strong angle/temperature-dependent Raman responses. *Nanoscale* **8**, 2063-2070 (2016).

20  Wang, H. & Qian, X. Giant optical second harmonic generation in two-dimensional multiferroics. *Nano Lett.* **17**, 5027-5034 (2017).

21  Patel, M., Chavda, A., Mukhopadhyay, I., Kim, J. & Ray, A. Nanostructured SnS with inherent anisotropic optical properties for high photoactivity. *Nanoscale* **8**, 2293-2303 (2016).

22    Baroni, S., De Gironcoli, S., Dal Corso, A. & Giannozzi, P. Phonons and related crystal properties from density-functional perturbation theory. *Rev. Mod. Phys.* **73**, 515 (2001).

23    Brent, J. R. *et al.* Tin (II) sulfide (SnS) nanosheets by liquid-phase exfoliation of herzenbergite: IV–VI main group two-dimensional atomic crystals. *J. Am. Chem. Soc.* **137**, 12689-12696 (2015).

24    Fuchs, G., Schiedel, C., Hangleiter, A., Härle, V. & Scholz, F. Auger recombination in strained and unstrained InGaAs/InGaAsP multiple quantum-well lasers. *Appl. Phys. Lett.* **62**, 396-398 (1993).

25    Klimov, V. I., Mikhailovsky, A. A., McBranch, D., Leatherdale, C. A. & Bawendi, M. G. Quantization of multiparticle Auger rates in semiconductor quantum dots. *Science* **287**, 1011-1013 (2000).

26    Wang, F., Dukovic, G., Knoesel, E., Brus, L. E. & Heinz, T. F. Observation of rapid Auger recombination in optically excited semiconducting carbon nanotubes. *Phys. Rev. B* **70**, 241403 (2004).

27    Sun, D. *et al.* Observation of rapid exciton–exciton annihilation in monolayer molybdenum disulfide. *Nano Lett.* **14**, 5625-5629 (2014).

28    Kumar, N. *et al.* Exciton-exciton annihilation in $MoSe_2$ monolayers. *Phys. Rev. B* **89**, 125427 (2014).

29    Yuan, L. & Huang, L. Exciton dynamics and annihilation in $WS_2$ 2D semiconductors. *Nanoscale* **7**, 7402-7408 (2015).

30    Wang, K. *et al.* Ultrafast saturable absorption of two-dimensional $MoS_2$ nanosheets. *ACS Nano* **7**, 9260-9267 (2013).

31   Lu, S. *et al.* Broadband nonlinear optical response in multi-layer black phosphorus: an emerging infrared and mid-infrared optical material. *Opt. Express* **23**, 11183-11194 (2015).

32   Lu, S. *et al.* Ultrafast nonlinear absorption and nonlinear refraction in few-layer oxidized black phosphorus. *Photonics Res.* **4**, 286-292 (2016).

33   Xu, Y. *et al.* Size-dependent nonlinear optical properties of black phosphorus nanosheets and their applications in ultrafast photonics. *J. Mater. Chem. C* **5**, 3007-3013 (2017).

34   Guo, Z. *et al.* From black phosphorus to phosphorene: basic solvent exfoliation, evolution of Raman scattering, and applications to ultrafast photonics. *Adv. Funct. Mater.* **25**, 6996-7002 (2015).

35   Wang, X. *et al.* Highly anisotropic and robust excitons in monolayer black phosphorus. *Nat. Nanotechnol.* **10**, 517 (2015).

36   Xu, L., Yang, M., Wang, S. J. & Feng, Y. P. Electronic and optical properties of the monolayer group-IV monochalcogenides M X (M= Ge, Sn; X= S, Se, Te). *Phys. Rev. B* **95**, 235434 (2017).

37   Persson, K. Materials Data on SnS (SG:62) by Materials Project.  (2014).

38   Kolobov, A., Fons, P. & Tominaga, J. Electronic excitation-induced semiconductor-to-metal transition in monolayer MoTe$_2$. *Phys. Rev. B* **94**, 094114 (2016).

39   Zhang, X., Tan, Q.-H., Wu, J.-B., Shi, W. & Tan, P.-H. Review on the Raman spectroscopy of different types of layered materials. *Nanoscale* **8**, 6435-6450 (2016).

40   Lin, S. *et al.* Accessing valley degree of freedom in bulk Tin (II) sulfide at room temperature. *Nat. Commun.* **9**, 1455 (2018).

41    Nikolic, P., Mihajlovic, P. & Lavrencic, B. Splitting and coupling of lattice modes in the layer compound SnS. *J.Phys.C: Solid State.Phys* **10**, L289 (1977).

42    Chandrasekhar, H., Humphreys, R., Zwick, U. & Cardona, M. Infrared and Raman spectra of the IV-VI compounds SnS and SnSe. *Phys. Rev. B* **15**, 2177 (1977).

43    Li, H. *et al.* From bulk to monolayer MoS$_2$: evolution of Raman scattering. *Adv. Funct. Mater.* **22**, 1385-1390 (2012).

44    Late, D. J., Maitra, U., Panchakarla, L., Waghmare, U. V. & Rao, C. Temperature effects on the Raman spectra of graphenes: dependence on the number of layers and doping. *J. Phys. Condens. Matter* **23**, 055303 (2011).

45    Sahoo, S., Gaur, A. P., Ahmadi, M., Guinel, M. J.-F. & Katiyar, R. S. Temperature-dependent Raman studies and thermal conductivity of few-layer MoS$_2$. *J. Phys. Chem. C* **117**, 9042-9047 (2013).

46    Abdula, D., Ozel, T., Kang, K., Cahill, D. G. & Shim, M. Environment-induced effects on the temperature dependence of Raman spectra of single-layer graphene. *J. Phys. Chem. C* **112**, 20131-20134 (2008).

47    Zouboulis, E. & Grimsditch, M. Raman scattering in diamond up to 1900 K. *Phys. Rev. B* **43**, 12490 (1991).

48    Sarkar, A. S. & Pal, S. K. Electron–Phonon Interaction in Organic/2D-Transition Metal Dichalcogenide Heterojunctions: A Temperature-Dependent Raman Spectroscopic Study. *ACS Omega* **2**, 4333-4340 (2017).

49    Menéndez, J. & Cardona, M. Temperature dependence of the first-order Raman scattering by phonons in Si, Ge, and α− S n: Anharmonic effects. *Phys. Rev. B* **29**, 2051 (1984).

50   Pawbake, A. S., Pawar, M. S., Jadkar, S. R. & Late, D. J. Large area chemical vapor deposition of monolayer transition metal dichalcogenides and their temperature dependent Raman spectroscopy studies. *Nanoscale* **8**, 3008-3018 (2016).

51   Sheik-Bahae, M., Said, A. A., Wei, T.-H., Hagan, D. J. & Van Stryland, E. W. Sensitive measurement of optical nonlinearities using a single beam. *IEEE J. Quantum Electron.* **26**, 760-769 (1990).

52   Tumuluri, A., Bharati, M., Rao, S. V. & Raju, K. J. Structural, optical and femtosecond third-order nonlinear optical properties of LiNbO3 thin films. *Mater. Res. Bull.* **94**, 342-351 (2017).

53   Li, P. *et al.* Two-dimensional $CH_3NH_3PbI_3$ perovskite nanosheets for ultrafast pulsed fiber lasers. *ACS Appl. Mater. Interfaces* **9**, 12759-12765 (2017).

54   McGehee, M. D. & Heeger, A. J. Semiconducting (conjugated) polymers as materials for solid-state lasers. *Adv. Mater.* **12**, 1655-1668 (2000).

55   Köhler, A., Wilson, J. S. & Friend, R. H. Fluorescence and phosphorescence in organic materials. *Adv. Mater.* **14**, 701-707 (2002).

56   Yuma, B. *et al.* Biexciton, single carrier, and trion generation dynamics in single-walled carbon nanotubes. *Phys. Rev. B* **87**, 205412 (2013).

57   Ghosh, S., Kushavah, D. & Pal, S. K. Unravelling the Role of Surface Traps on Carrier Relaxation and Transfer Dynamics in Ultrasmall Semiconductor Nanocrystals. *J. Phys. Chem. C* (2018).

58   Kumar, P., Kumar, S., Ghosh, S. & Pal, S. K. Femtosecond insights into direct electron injection in dye anchored ZnO QDs following charge transfer excitation. *Phys. Chem. Chem. Phys.* **18**, 20672-20681 (2016).

*Supplementary Information*

**Strong many-body interactions in ultrathin anisotropic tin (II) monosulfide**

Abdus Salam Sarkar, Aamir Mushtaq, Dushyant Kushavah and Suman Kalyan Pal*

School of Basic Sciences, Indian Institute of Technology Mandi, Kamand, Mandi-175005, Himachal Pradesh, India.

Advanced Materials Research Centre, Indian Institute of Technology Mandi, Kamand, Mandi-175005, Himachal Pradesh, India.

Energy Center, Indian Institute of Technology Mandi, Kamand, Mandi-175005, Himachal Pradesh, India.

*Corresponding author
Email: suman@iitmandi.ac.in

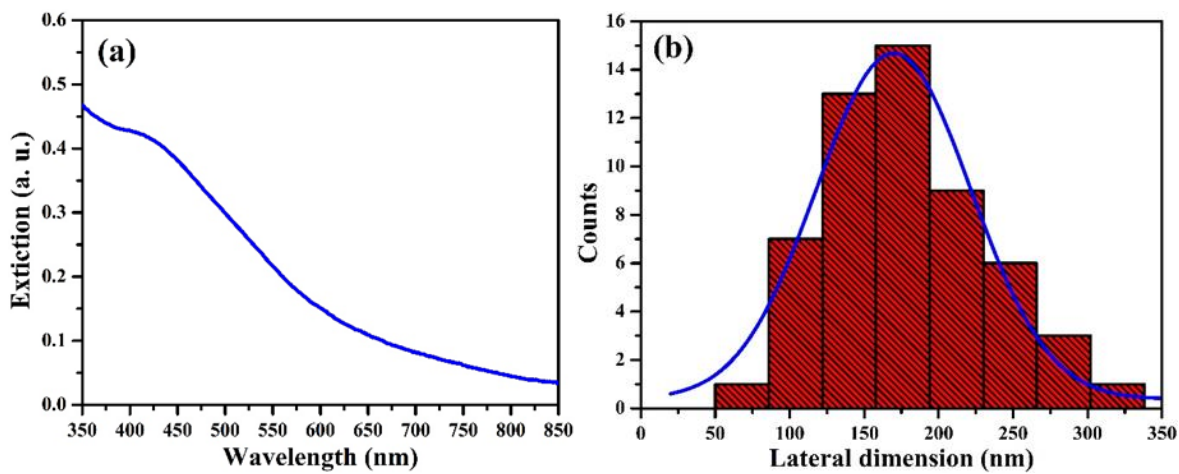

**Figure 1 | Optical and structural properties of exfoliated SnS nanosheets in acetone. a**, Extinction spectra and **b**, histogram of lateral dimension.

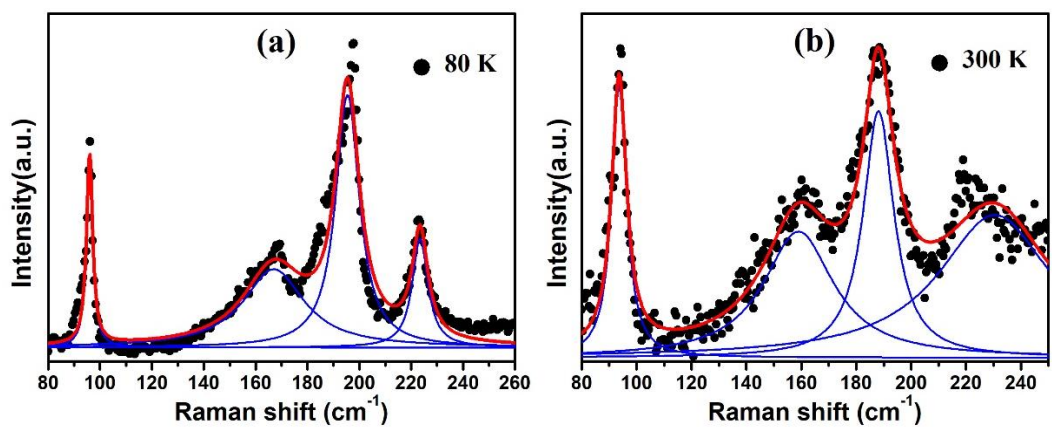

**Figure 2 | Raman shift of various vibrational modes of SnS nanosheets at 80 and 300 K.** Experimental data were fitted with Lorentz function. Solid lines are fitting results.

**Table 1 | Raman behavior of ultrathin SnS nanosheets in comparison to other 2D materials**

| Material | $\lambda_{laser}$ (nm) | Raman modes | $\chi$ (cm$^{-1}$/K) | Reference |
|---|---|---|---|---|
| Ultrathin SnS (1.10 nm thick) | 532 | $A_g$ (93.6 cm$^{-1}$) | -0.0105±0.0009 | This work |
| | | $B_{3g}$ (161.2 cm$^{-1}$) | -0.0280±0.0034 | |
| | | $A_g$ (187.8 cm$^{-1}$) | -0.0466±0.0047 | |
| | | $A_g$ (220.9 cm$^{-1}$) | -0.0129±0.0020 | |
| SnS flake (14.6 nm thick) | 532 | $A_g$ (95.5 cm$^{-1}$) | -0.016 | 1 |
| | | $B_{3g}$ (162.9 cm$^{-1}$) | -0.029 | |
| | | $A_g$ (190.7 cm$^{-1}$) | -0.036 | |
| | | $A_g$ (216.7 cm$^{-1}$) | -0.023 | |
| SnSe thin flake | 532 | $B_{3g}$ | -0.0331 | 2 |
| | | $A_g^2$ | -0.0377 | |
| | | $A_g^3$ | -0.0153 | |
| Few-layer black phosphorus (6 nm thick) | 532 | $A_g^1$ | -0.0164 | 3 |
| | | $B_{2g}$ | -0.0271 | |
| | | $A_g^2$ | -0.0283 | |
| Monolayer MoS$_2$ | 514.5 | $E_{2g}^1$ | -0.013 | 4 |
| | | $A_g^1$ | -0.011 | |
| Few-layer MoS$_2$ | 514.5 | $E_{2g}^1$ | -0.016 | 5 |
| | | $A_g^1$ | -0.008 | |
| Single layer graphene | 632.8 | G | -0.016 | 6 |
| | | 2D | -0.026 | |

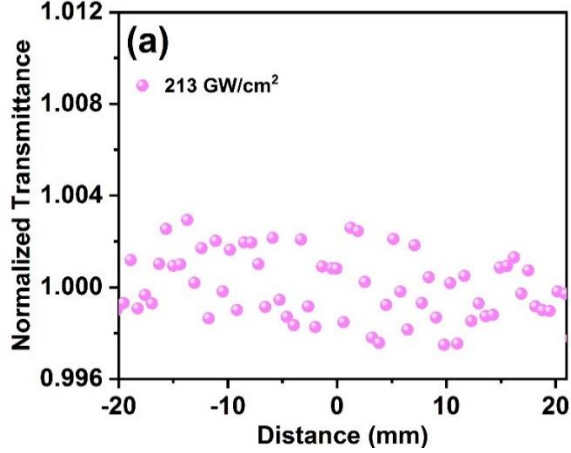

**Figure 3 | Z-scan results of the acetone solvent for the fs pulses at 800 nm.**

**Z-scan data analysis**

Normalized transmittance can be expressed as

$$T = \frac{1}{\sqrt{\pi q_0}} \int_{-\infty}^{\infty} \ln(1 + q_0 e^{-x^2}) dx \quad \ldots\ldots(1)$$

where $q_0 = \alpha_{NL} I_z L_{eff}$, $\alpha_{NL}$ is NLO absorption cofficient, $L_{eff} = \frac{(1-e^{-\alpha_0 L})}{\alpha_0 L}$, effective path length, $\alpha_0$ is linear absorption coefficient, L is the sample path length, $I_z = \frac{I_0}{1+\left(\frac{Z}{Z_0}\right)^2}$ where $I_0$ is on-axis peak intensity and $Z_0$ is the diffraction length of the beam. For $|q_0| < 1$, the transmittance in terms of peak irradiance can be expressed in summation form as[7,8]

$$T = \sum_{m=0}^{\infty} \frac{[-q_0(Z,0)]^m}{(m+1)^{\frac{3}{2}}} \quad \ldots\ldots(2)$$

where m is an integer. For m =1 equation (2) becomes

$$T = 1 - \frac{\alpha_{NL} I_0 L_{eff}}{2\sqrt{2}\left[1+\left(\frac{Z}{Z_0}\right)^2\right]} \quad \ldots\ldots(3)$$

This equation was used for fitting the open aperture Z-scan data.

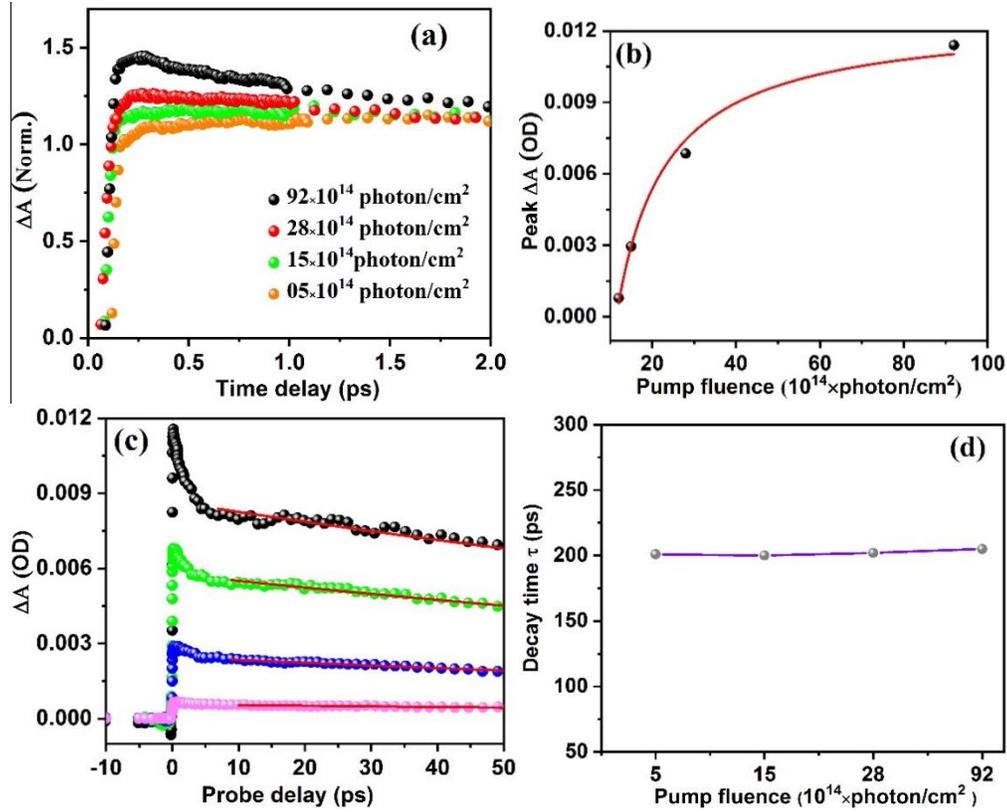

**Figure 4 | Transient absorption study of ultrathin SnS nanosheets. a,** Normalized TA kinetics at different pump fluences (pump and prove wavelengths are 480 and 550 nm, respectively. Here, each TA kinetics is divided by its DA value at 6 ps for clear understanding of the difference at early time. **b,** Peak TA signal as a function of injected exciton density. **c,** TA kinetics at 550 nm for various pump intensities: $92\times 10^{14}$ photons/cm$^2$ (black), $28\times 10^{14}$ photons/cm$^2$ (green), $15\times 10^{14}$ photons/cm$^2$ (blue) and $5\times 10^{14}$ photons/cm$^2$ (pink). Solid lines are fitting of the signal after 10 ps using a single exponential function having decay time t. **d,** Plot of t vs pump fluence. Straight line (parallel to x-axis) nature of this plot suggests that the decay kinetics after 10 ps is independent of incident photon density.

**Table 2 | Comparison of NLO properties of different 2D materials**

| Wavelength (nm) | Sample (Thickness, Size) | $\alpha_{NL}$ $\left(\frac{cm}{GW}\right)$ | $I_{sat}$ $\left(\frac{GW}{cm^2}\right)$ | Reference |
|---|---|---|---|---|
| 800 | SnS (1.1 nm, 180 nm) | $-0.5 \times 10^{-3}$ | 570 | This work |
| 800 | BP (5-10 nm, several 100 nm) | $-4.08 \times 10^{-3}$ | 647.7 | 9 |
| 800 | Phosphorene (2.8 nm, unknown) | _ | 774.4 | 10 |
| 800 | $MoS_2$ (4 nm, several 100 nm) | $-4.60 \times 10^{-3}$ | 413 | 11 |
| 800 | $MoSe_2$ (few nm, unknown) | $-2.54 \times 10^{-3}$ | 590 | 12 |
| 800 | Graphene (unknown, unknown) | $-8.28 \times 10^{-3}$ | 764 | 13 |

**References**

1   Xia, J. *et al.* Physical vapor deposition synthesis of two-dimensional orthorhombic SnS flakes with strong angle/temperature-dependent Raman responses. *Nanoscale* **8**, 2063-2070 (2016).

2   Luo, S. *et al.* Temperature-Dependent Raman Responses of the Vapor-Deposited Tin Selenide Ultrathin Flakes. *J. Phys. Chem. C* **121**, 4674-4679 (2017).

3   Łapińska, A., Taube, A., Judek, J. & Zdrojek, M. Temperature evolution of phonon properties in few-layer black phosphorus. *J. Phys. Chem. C* **120**, 5265-5270 (2016).

4   Pawbake, A. S., Pawar, M. S., Jadkar, S. R. & Late, D. J. Large area chemical vapor deposition of monolayer transition metal dichalcogenides and their temperature dependent Raman spectroscopy studies. *Nanoscale* **8**, 3008-3018 (2016).

5   Thripuranthaka, M., Kashid, R. V., Sekhar Rout, C. & Late, D. J. Temperature dependent Raman spectroscopy of chemically derived few layer MoS2 and WS2 nanosheets. *Appl. Phys. Lett.* **104**, 081911 (2014).

6   Late, D. J., Maitra, U., Panchakarla, L., Waghmare, U. V. & Rao, C. Temperature effects on the Raman spectra of graphenes: dependence on the number of layers and doping. *J. Phys. Condens. Matter* **23**, 055303 (2011).

7   Sheik-Bahae, M., Said, A. A., Wei, T.-H., Hagan, D. J. & Van Stryland, E. W. Sensitive measurement of optical nonlinearities using a single beam. *IEEE J. Quantum Electron.* **26**, 760-769 (1990).

8   Li, P. *et al.* Two-dimensional $CH_3NH_3PbI_3$ perovskite nanosheets for ultrafast pulsed fiber lasers. *ACS Appl. Mater. Interfaces* **9**, 12759-12765 (2017).

9   Lu, S. *et al.* Broadband nonlinear optical response in multi-layer black phosphorus: an emerging infrared and mid-infrared optical material. *Opt. Express* **23**, 11183-11194 (2015).

10  Guo, Z. *et al.* From black phosphorus to phosphorene: basic solvent exfoliation, evolution of Raman scattering, and applications to ultrafast photonics. *Adv. Funct. Mater.* **25**, 6996-7002 (2015).

11      Wang, K. *et al.* Ultrafast saturable absorption of two-dimensional MoS$_2$ nanosheets. *ACS Nano* **7**, 9260-9267 (2013).

12      Wang, K. *et al.* Broadband ultrafast nonlinear absorption and nonlinear refraction of layered molybdenum dichalcogenide semiconductors. *Nanoscale* **6**, 10530-10535 (2014).

13      Wang, K. *et al.* Ultrafast nonlinear excitation dynamics of black phosphorus nanosheets from visible to mid-infrared. *ACS Nano* **10**, 6923-6932 (2016).